\documentstyle[aps,prl,preprint,epsfig]{revtex}
\begin{document}
\draft

\preprint{\tighten\vbox{\hbox{\hfil BIHEP-EP1-99-07}}}
\title{Partial Wave Analysis of $J/\psi\to\gamma(\pi^+\pi^-\pi^+\pi^-)$}

\author{
J.~Z.~Bai,$^{1}$    Y.~Ban,$^{6}$      J.~G.~Bian,$^{1}$  A.~D.~Chen,$^{1}$
G.~P.~Chen,$^{1}$   H.~F.~Chen,$^{2}$  H.~S.~Chen,$^{1}$  J.~C.~Chen,$^{1}$
X.~D.~Chen,$^{1}$   Y.~Chen,$^{1}$
% ...
\\(BES Collaboration)\cite{besjpsi}\\
D.~V.~Bugg,$^{12}$  B.~S.~Zou $^{1,12}$ and I.~Scott $^{12}$}

\vspace{1cm}

\address{
$^{1}$ Institute of High Energy Physics, Beijing 100039,
People's Republic of China \\
$^{2}$ University of Science and Technology of China, Hefei 230026,
People's Republic of China \\
$^{3}$ Shandong University, Jinan 250100,
People's Republic of China \\
$^{4}$ Hangzhou University, Hangzhou 310028,
People's Republic of China \\
$^{5}$ Shanghai Jiaotong University, Shanghai 200030,
People's Republic of China \\
$^{6}$ Peking University, Beijing 100871,
People's Republic of China \\
$^{7}$ Hua Zhong Normal University, Wuhan 430079,
People's Republic of China \\
$^{8}$ China Center for Advanced Science and Technology(CCAST), World
Laboratory, Beijing 100080, People's Republic of China)\\
$^{9}$ Nankai University, Tianjin 300071,
People's Republic of China \\
$^{10}$ Henan Normal University, Xinxiang 453002,
People's Republic of China \\
$^{11}$ Wuhan University, Wuhan 430072,
People's Republic of China \\
$^{12}$ Queen Mary and Westfield College, London E1 4NS, United Kingdom}


\maketitle

\begin{abstract}
BES data on $J/\psi \to \gamma (\pi^{+} \pi^{-} \pi^{+} \pi^{-})$
have been analyzed into partial waves.
We fit with resonances having $J^{PC}=2^{++}$ at 1275 MeV,
$0^{++}$ at 1500 MeV, $2^{++}$ at 1565 MeV,
$0^{++}$ at 1740 MeV, $2^{++}$ at 1940 MeV and $0^{++}$ at 2104 MeV,
plus a broad $0^-$ component.
The $0^{++}$ resonances decay dominantly to $\sigma\sigma$,
while $2^{++}$ resonances in the high mass region decay mainly to
$f_2(1270)\sigma$ and $\sigma\sigma$; $2^{++}$ resonances from the low 
mass region decay dominantly to $\rho\rho$.
\end{abstract}

\vspace{0.6cm}

MARK III and DM2 reported their data on
$J/\psi \to \gamma (\pi^{+}\pi^{-}\pi^{+}\pi^{-})$ in 1986 and 1989
respectively \cite{mark3,dm2}. They found that the dominant component
of the spectrum below 2 GeV was due to $J/\psi\to\gamma\rho\rho$
with spin-parity $0^-$ in the $\rho\rho$ system, where the two $\rho$
appear in a relative P-wave. MARK III claimed a pseudoscalar resonance
at $1.55$ GeV/c$^2$. This was supported by DM2, and in addition they
claimed that structures at $1.80$ and $2.10$ GeV/c$^2$ were also
$J^{PC}=0^{-+}$.

Later, E760 found three structures in $\eta\eta$ \cite{e760} with
masses and widths very similar to those observed in
$J/\psi \to \gamma \pi^{+}\pi^{-}\pi^{+}\pi^{-}$. 
However, the quantum numbers $0^{-+}$ are forbidden to $\eta\eta$.
Both MARK III and DM2 analyzed their data in terms of
$J/\psi \to \gamma\rho\rho \to \gamma\pi^{+}\pi^{-}\pi^{+}\pi^{-}$,
but they did not consider
$J/\psi \to \gamma\sigma\sigma \to \gamma\pi^{+}\pi^{-}\pi^{+}\pi^{-}$. 
Here $\sigma$ means the full $\pi -\pi$ S-wave amplitude \cite{zou},
parametrized up to 1800 MeV.
In 1995, Bugg {\it {et al.}} reanalyzed MARK III data \cite{bugg} and
added the $\sigma\sigma$ decay mode to $\rho\rho$. Inclusion of this
decay identified two of the peaks as $I=0$ scalar resonances decaying
exclusively via $\sigma\sigma$. Those states have masses in the region
$M=1500$ and $2100$ MeV/c$^2$. An additional scalar state was required
at $M=1750$ MeV/c$^2$, decaying dominantly to $\sigma\sigma$, but also
with significant decays via $\rho\rho$.

We present here an analysis of BES data on
$J/\psi\to\gamma\pi^{+}\pi^{-}\pi^{+}\pi^{-}$ into partial waves.
In this analysis, the full Monte Carlo simulation of the BES detector
is used. This improves on the analysis of Bugg {\it {et al.}}, which 
 used only a simplified
Monte Carlo simulation, since the full Monte Carlo simulation of the
Mark III detector was then no longer available.
The full Monte Carlo simulation allows an improved study of the
main background channel
$J/\psi\to\pi^0\pi^{+}\pi^{-}\pi^{+}\pi^{-}$.
Results are similar to those of Ref. \cite{bugg}, but with the significant 
addition that a 
$2^+$ amplitude is required in the high mass region around 2 GeV.

The Beijing Spectrometer(BES) has collected $7.8 \times
10^6 ~J/\psi$ triggers, used here.
Details of the detector are given in Ref. \cite{detect}.
We describe briefly those detector elements playing a crucial role in
the present measurement.
Tracking is provided by a 10 superlayer main drift chamber (MDC).
Each superlayer contains four layers of sense wires measuring both the
position and the ionization energy loss ($d$E$/dx$) of charged particles.
The momentum resolution is $\sigma_P/P = 1.7\%\sqrt{1 + P^2}$,
where $P$ is the momentum of charged particles in GeV/$c$.
The resolution of the $d$E$/dx$ measurement is $\sim \pm 9\%$,
providing good $\pi/K$ separation and proton identification for momenta
up to 600 MeV/c.
An array of 48 scintillation counters surrounding the MDC measures the
time-of-flight (TOF) of charged particles with a resolution of 330$ps$ for
hadrons. Outside the TOF system is an electromagnetic calorimeter made
of lead sheets and streamer tubes and having a $z$ positional resolution
of 4 cm. The energy resolution scales as $\sigma_E/E = 22\%/\sqrt{E}$, 
where $E$ is the energy in GeV. Outside the shower counter is a solenoidal
magnet producing a 0.4 Tesla magnetic field.

Candidates for the decay $J/\psi\to\gamma\pi^{+}\pi^{-}\pi^{+}\pi^{-}$
are selected by requiring exactly four charged tracks. Every track must have
a good helix fit in the polar angle range $-0.8 < \cos\theta < 0.8$ and
a transverse momentum $>60$ MeV/c.
A vertex is required within an interaction region
$\pm 20$ cm longitudinally and 2 cm radially. 
The number of neutral
clusters may be up to five, but only one reconstructed $\gamma$ is required
in the barrel shower counter. A minimum energy cut of 50 MeV is imposed
on the photons. 
Showers that can be associated with charged tracks are not considered.
Four-constraint kinematic fits are performed to the final states
$\gamma\gamma 4\pi$, $\gamma4\pi$ and $4\pi$. The value of
$\chi^2(\gamma4\pi)$ is required to be the smallest one. 
The probability of $\chi^2(\gamma4\pi)$ is required to be larger than
$5$\%, in order to achieve good resolution and to select a good photon.

Further cuts are as follows.
To remove the main background, 
$J/\psi\to\pi^0\pi^{+}\pi^{-}\pi^{+}\pi^{-}$,
we required the probability $Prob(\chi^2(\gamma4\pi))$ to be larger than
$Prob(\chi^2(\gamma\gamma_{miss}4\pi))$,
where $\gamma_{miss}$ means this photon is missing.
If the two photons in $J/\psi\to\gamma\gamma_{miss}
\pi^{+}\pi^{-}\pi^{+}\pi^{-}$ are from a $\pi^0$ decay,
$Prob(\chi^2(\gamma\gamma_{miss}4\pi))$ is required to be $<1$\%. Next,
$\mid{U_{miss}}\mid = \mid E_{miss} - P_{miss} \mid < 0.12$ GeV/c$^2$
is required in order to reject events containing more than one photon
or containing charged kaons; here, $E_{miss}$ and $P_{miss}$ are,
respectively, the missing energy and missing momentum of all charged
particles, which are taken to be $\pi^{+}\pi^{-}\pi^{+}\pi^{-}$.
The transverse momentum of the $4\pi$ system
$P_{t\gamma}^2 = 4\mid{P_{miss}}\mid{^2}~\sin^2(\theta_{m\gamma}/2)$
is required to be                   
$<0.005$ (GeV/c)$^2$, in order to remove the background
$J/\psi\to\pi^0\pi^{+}\pi^{-}\pi^{+}\pi^{-}$; here $\theta_{m\gamma}$ is
the angle between the missing momentum and the photon direction.
Background $J/\psi\to\omega\pi^+\pi^-$ events are eliminated by the cut
$\mid M_{\pi^+\pi^-\pi^0}-M_{\omega} \mid > 25$ MeV, in the
$\pi^{0}\pi^{+}\pi^{-}\pi^{+}\pi^{-}$ hypothesis with only one photon
detected and the $\pi^{0}$ associated to the missing momentum.
To remove the small background due to $J/\psi\to\gamma K_s^0K_s^0$ events,
a double cut on the two $\pi^+\pi^-$ invariant masses is performed,
$\mid M_{\pi^+\pi^-}-M_{K_s^0} \mid >25$ MeV.
The BES data show a small signal due to  $f_1(1285)$; this is not
important for our conclusion and not
of concern in this analysis, so, in order to remove it, an additional
cut is applied to discard events within the mass region between
$1.24$ and $1.32$ GeV/c$^2$.

The effects of the various selection cuts on the data is simulated with
a full Monte Carlo of the BES detector; 200,000 Monte Carlo events are
successfully fitted to $J/\psi \to \gamma (\pi^+\pi^-\pi^+\pi^-)$
with $4\pi$ mass below 2.4 GeV;
all background reactions are similarly fitted to this channel. 
The estimated background is 31\%, purely from
$J/\psi \to \pi ^0 (\pi^+\pi^-\pi ^+\pi ^-)$. We have included this
background in the amplitude analysis;  it lies very close to 
phase space of $J/\psi \to \gamma (\pi^+\pi^-\pi^+\pi^-)$ and has
only a small effect on the fit.
It includes a small signal (visible in Fig.~\ref{figure3}(c) and (d) below)
for $a_2(1320) \to \rho \pi$, originating from
$J/\psi \to a_2\rho$. When this $a_2(1320)$ signal is combined with a fourth
pion, it does not correlate with any peak in the $4\pi$ mass spectrum.
Otherwise, the $4\pi$ and $3\pi$ mass distributions in the background
lie very close to phase space and are parametrised by phase space
distributions. 

Fig.~\ref{figure1} shows the $2\pi$ invariant mass distribution after 
acceptance cuts. The acceptance from the Monte Carlo simulation is included
into the maximum likelihood fit. Signals are seen due to $\rho$ (or $\sigma$)
and $f_2(1270)$.
The $\sigma$ contains two components, one very broad and the other
with a peak near the $\rho$ and somewhat wider than the $\rho$. 
It cannot be reliably separated from the $\rho$ as a peak. The two are
distinguished by their quite different angular dependence.
They are quite easy to distinguish in the fit, but this is not easily
displayed graphically, because of the two combinations.
The lowest peak of Fig.~\ref{figure1}(b) is due to
conversion electrons, and has been cut out in the fit.
The $4\pi$ invariant mass distribution for 
$J/\psi\to\gamma\pi^{+}\pi^{-}\pi^{+}\pi^{-}$
is shown with error bars in Fig.~\ref{figure2}; the amplitude analysis resolves
structures from $1.50$ to $2.30$ GeV/c$^2$.
The dark dashed histogram of Fig.~\ref{figure2} shows the $4\pi$ mass
spectrum of $J/\psi\to\gamma f_2(1270)\sigma$, which rises strongly 
above 1750 MeV. 
For our final fit, we use 1988 events below a $4\pi$ mass of 2.4 GeV.

We have carried out a partial wave analysis using amplitudes constructed
from Lorentz-invariant combinations of the 4-vectors and the photon
polarization for $J/\psi$ initial states with helicity $\pm 1$. 
Cross sections are summed over photon polarizations. The relative
magnitudes and phases of the amplitudes are determined by a maximum
likelihood fit. We use $\ell$ to denote the orbital angular momentum between
the photon from $J/\psi$ decay and the resonance in the production process;
we use $L$ to denote the orbital angular momentum in the decay of the resonance
to final states $\rho \rho$, $\sigma \sigma$ and $f_2(1270)\sigma$, with $L$
restricted $\le 2$. 
Because production is via an electromagnetic
transition, the same phase is used for amplitudes with different $\ell$
but otherwise the same final state. 

We now discuss resonances and possible decay modes which have been
included in the fit. 
Spin-parity assignments up to $J = 4$ have been tried.  
We have tried to include $\pi - \pi(1300)$ or $\pi - a_1(1260)$
decay modes for $0^+$ resonance. There are two decay modes of 
$\pi(1300)$ and $a_1(1260)$: $\rho\pi$ and $\sigma\pi$.
From the PDG \cite{pdg1}, $\pi(1300)$ and $a_1(1260)$ decay mainly to 
$\rho\pi$. When putting this constraint into the fit, we found  
the contributions of the $\pi - \pi(1300)$ and $\pi - a_1(1260)$ decay
modes are negligible. The resonances and decay modes
which make significant improvements in log likelihood are listed in
Table~\ref{table1}.  Amplitudes are constructed in terms of the
combined spin $S$ of the decay photon and the $J/\psi$ system.

The dominant component in the fit is $0^-$. We find that this may be
parametrized in two alternative ways.
The one used for the fit reported here is a simple Breit-Wigner
amplitude of mass 1440 MeV and width 225 MeV. It is much wider than
$\eta (1440)$ and is therefore not to be identified with that state.
The rapidly increasing phase space for decays to $\rho \rho$ with
$L = 1$ results in a very broad $0^-$ signal, illustrated below in
Fig.~\ref{figure4}(a). An alternative more complicated parametrization
is given in a recent coupled channel analysis of many $J/\psi$ decay
modes \cite {coupled};
it gives results almost indistinguishable from those presented here.

The width of $f_2(1270)$ cannot be fitted with any precision.
Statistics of data in the $f_2(1270)$ mass region are very low because
the $f_1(1285)$ has been removed by a cut in this mass region.
Hence the mass and width of the $f_2(1270)$ are constrained to the
PDG value~\cite{pdg1}.
Some $2^+$ component is definitely required in the mass range
1500--1700 MeV, as is illustrated below in Fig.~\ref{figure4}(c). 
This may be attributed to $f_2(1565)$, which sits at the 
$\rho\rho$ and $\omega\omega$ threshold and can decay into $4\pi$. 
In the final fit, the mass of $f_2(1565)$ has been fixed at 1565 MeV,
which is well determined by data from other sources \cite {May}
\cite {Abele} \cite {Hodd}. The possible
cause of our optimum mass for $f_2(1565)$ lying at 1505 MeV may be
crosstalk between the two neighboring resonances
($f_0(1500)$ and $f_2(1565)$). Once the masses and widths are fixed,
both contributions are well determined.

Comparisons with data are shown in Fig.~\ref{figure3} summed from
$M(4\pi) = 1.0$ to 2.4 GeV for
$M_{\pi^+\pi^-\pi^+\pi^-}$, $M_{\pi^+\pi^-}$, $M_{3\pi}$, $M_{\rho\pi}$, 
$\cos \theta _{\pi^+}$; here $\theta _{\pi^+}$ is the angle of $\pi^{+}$
with respect to the $\pi^{+}\pi^{-}$ pair in their rest frame, 
and $\chi$ is the azimuthal angle between the planes of $\pi^{+}\pi^{-}$
pairs in the rest frame of the resonance X.
The contributions of the various components in this fit are shown in
Fig.~\ref{figure4} and Fig.~\ref{figure5};
crosses are data and histograms the fit.

From Fig.~\ref{figure5}, it can be seen that $2^+$ resonances in the
high mass region decay mainly via $f_2(1270)\sigma$ and $\sigma\sigma$;
the $2^+$ resonances in the low mass region decay mainly via $\rho\rho$.
The peak of $f_2(1565)$ appears at $\sim 1580$ MeV because of the rapidly
increasing phase space for its $4\pi$ decay.
The $2^+$ signals at about 1580 MeV with L=0 and L=2 are both larger than
the combined signals in Figure~\ref{figure4}; the reason for this is
destructive interference between L=0 and L=2.

Branching fractions are given in Table~\ref{table2}. 
The branching fraction for $M_{4\pi} < 3.0$ GeV/c$^2$ is
$Br(J/\psi\to\gamma\pi^{+}\pi^{-}\pi^{+}\pi^{-})
=(3.8\pm0.3\pm1.3)\times10^{-3}.$ The first error is statistical and
the second systematic.

In order to check whether the final full fit is the best solution
for BES data, four tests are performed.

The first test is to check the significance level of each resonance.
The $0^-$ contribution of Fig.~\ref{figure4} is definitely required;
without it, log likelihood gets worse by 293.1.
Table~\ref{table3}  summarizes the changes in log likelihood when each
resonance of the fit is dropped and remaining contributions are
re-optimized. It indicates that the $f_2(1270)$, $f_0(1500)$,
$f_2(1565)$, $f_0(1740)$, $f_2(1950)$ and $f_0(2100)$ should be included.

The second test is to check the $J^P$ assignment of each resonance.
Table~\ref{table4} shows the effect of substituting alternative $J^P$
assignments for each resonance. 
It can be seen that making an alternative $J^P$ assignment for each
resonance is rejected clearly.

The third test is to scan for an additional structure.
Scanning an additional structure in $0^-$, $0^+$, $2^+$, or $4^+$
with 2, 4, 6 and 4 more free parameters respectively
indicates no further distinctive structures.
The BES data on $J/\psi \to\gamma K^+K^-$~\cite{bes1} show two possible
resonance in the 1.7 GeV region: a scalar with mass 1781 MeV,
a tensor 1696 MeV. We have therefore tried adding to the fit a 
tensor resonance in this mass region. We have found no significant
production of spin 2 is observed in the $\gamma \pi^+\pi^-\pi^+\pi^-$
channel.

The fourth test is to fit the data in slices of mass.
The data are fitted in each of 10 mass bins 100 MeV wide
from 1350 to 2350 MeV.
The comparisons of the full fit and the slice fit are plotted in
Fig.~\ref{figure6}.
The solid line presents the full fit results, while the dot stands for
the slice analysis results. They are in fair agreement with each other.
The slice analysis confirms strongly the presence of a tensor component
in the mass region around 2000 MeV.

The results are summarized as follows:
The $0^{++}$ resonances decay dominantly to $\sigma\sigma$, while
the $2^{++}$ resonances in the low mass region decay dominantly
to $\rho\rho$,  and those in the high mass region decay to
$f_2(1270)\sigma$, $\sigma\sigma$.

The $f_0(1500)$ and $f_0(1740)$ are produced significantly in
$J/\psi\to\gamma\pi^{+}\pi^{-}\pi^{+}\pi^{-}$; according to the criteria
of determining the gluonic content of resonance in radiative $J/\psi$
decays proposed by Close {\it {et al.}}\cite{close},
they may therefore contain significant glueball components.

The BES results are approximately consistent with the results from the
earlier re-analysis of MARK III data [4]. But the BES data favour more
$2^+$ in the high mass region. A possible cause of this difference is
that a better background function for BES data is available and a full
Monte Carlo simulation is employed in BES analysis. 
An important conclusion is that a broad $2^+$ resonance $f_2(1950)$ is
needed around 2 GeV. This is where the $2^+$ glueball may be expected,
and $J/\psi$ radiative decays are supposed to be one of the best places
to search for it. The $f_2(1950)$ was first observed by the 
WA91 group~\cite{wa91} and subsequently confirmed by the WA102 group
in their $4\pi$ mass spectrum~\cite{wa102};
this agrees with our observation.

{\vspace{0.9cm}
The BES group thanks the staff of IHEP for technical support in running the
experiment. This work is supported in part by China Postdoctoral Science
Foundation and National Natural Science Foundation of China under contract
Nos. 19991480, 19825116 and 19605007;
and by the Chinese Academy of Sciences under contract No. KJ 95T--03(IHEP).
We also acknowledge financial support from the Royal Society
for collaboration between Chinese and UK groups.

\begin{center}

\begin{figure}[htbp]
\centerline{\epsfig{file=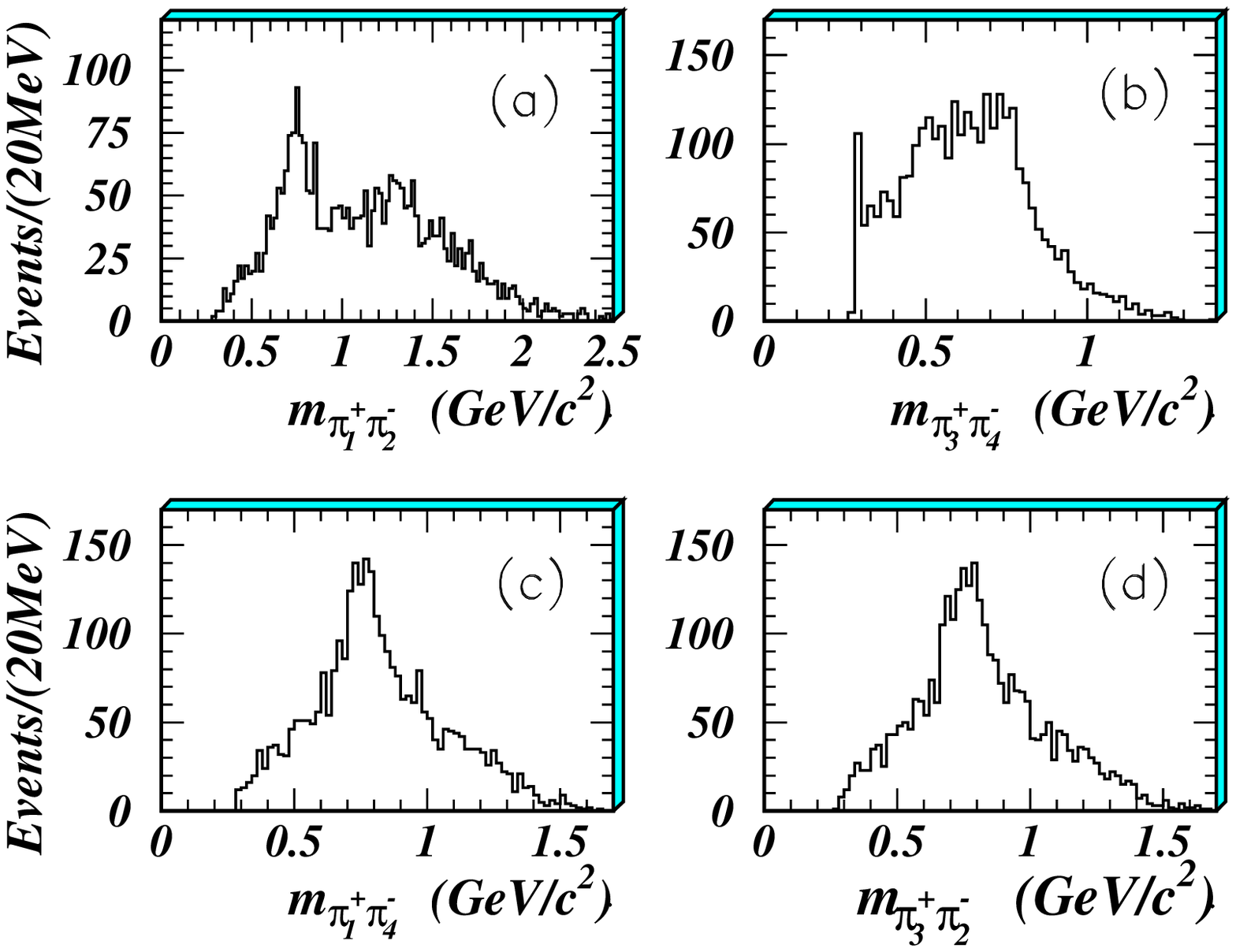,width=5.0in}} 
\caption[]{The $2\pi$ mass spectrum of
$J/\psi\to\gamma\pi^+\pi^-\pi^+\pi^-$,
taking $P(\pi_1^+) > P(\pi_3^+)$ and  $P(\pi_2^-) > P(\pi_4^-)$.}
\label{figure1}
\end{figure}

\begin{figure}[htbp]
\centerline{\epsfig{file=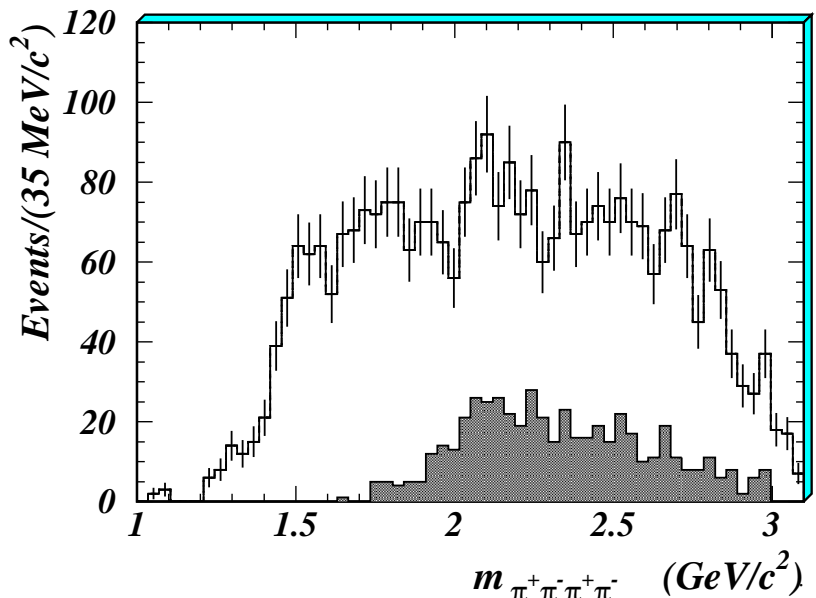,width=4.5in}} 
\caption[]{The $4\pi$ mass spectrum of
$J/\psi\to\gamma\pi^+\pi^-\pi^+\pi^-$; the dark shaded histogram
is the $4\pi$ mass spectrum of $J/\psi\to\gamma f_2(1270) \sigma$}
\label{figure2}
\end{figure}

\begin{figure}[htbp]
\centerline{\epsfig{file=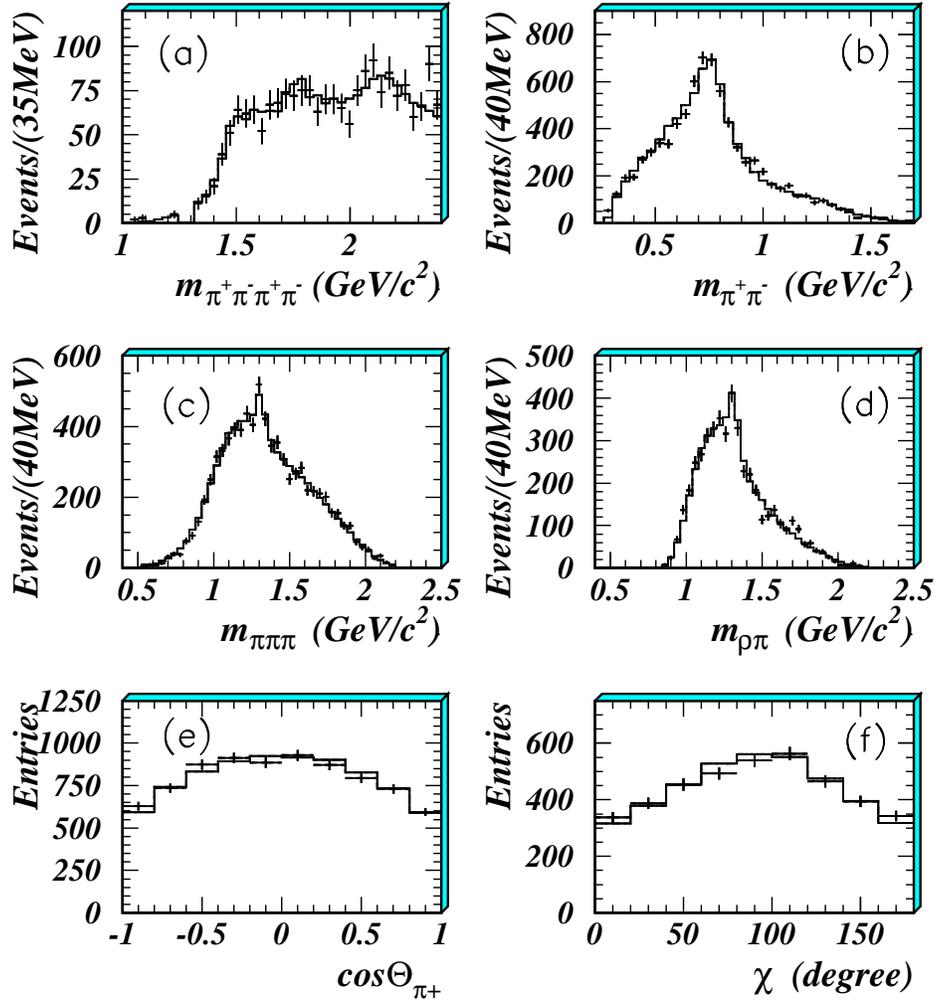,width=5.0in}} 
\caption[]{The comparison between data and final full fit.}
\label{figure3}
\end{figure}

\begin{figure}[htbp]
\centerline{\epsfig{file=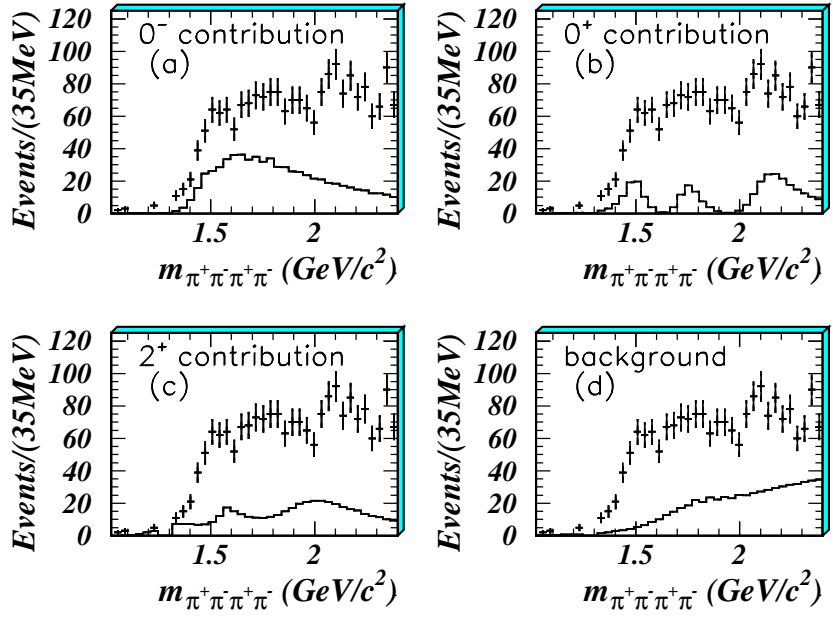,width=4.5in}}
\caption[]{component contribution.}
\label{figure4}
\end{figure}

\begin{figure}[htbp]
\centerline{\epsfig{file=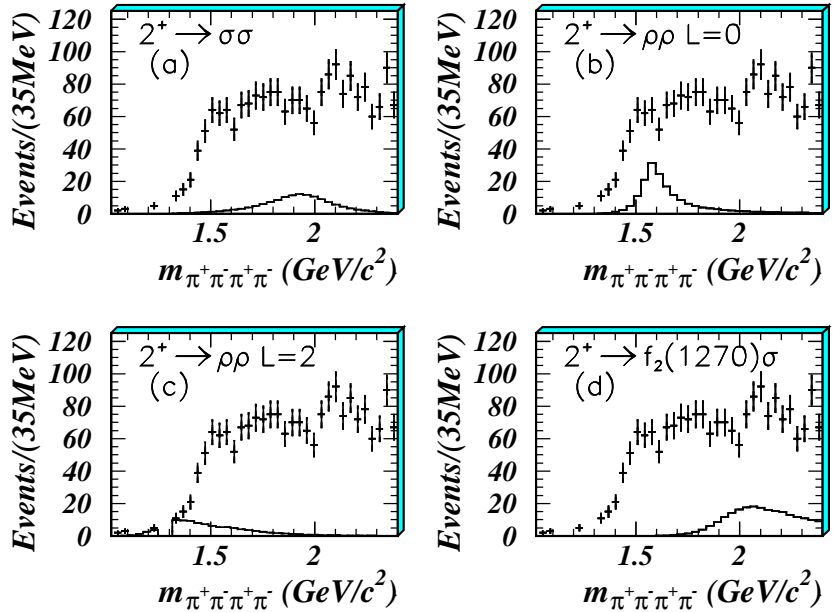,width=4.5in}} 
\caption[]{tensor component contribution.}
\label{figure5}
\end{figure}

\begin{figure}[htbp]
\centerline{\epsfig{file=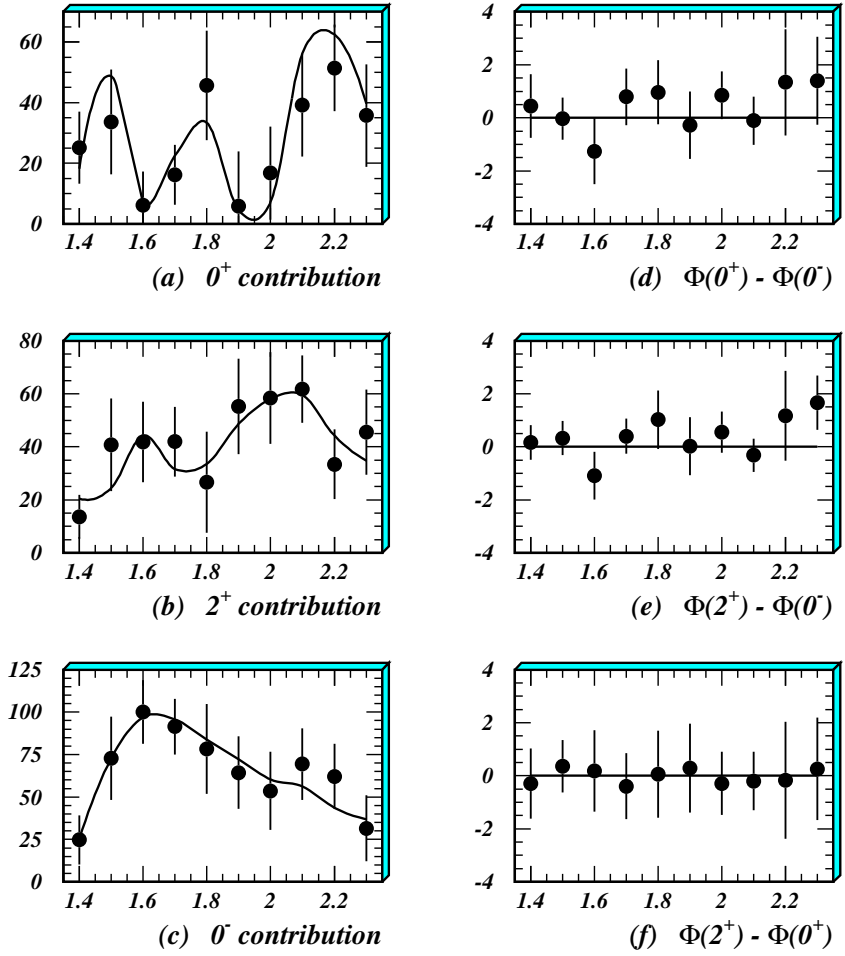,width=5.0in}} 
\caption[]{Numbers of events in 100 MeV slices of $M(4\pi)$ (data point)
from (a) $0^+$, (b) $2^+$ and (c) $0^-$ and the full fit (curves);
the discrepancy in phase difference (d) $\phi(0^+)-\phi(0^-)$,
(e) $\phi(2^+)-\phi(0^-)$ and (f) $\phi(2^+)-\phi(0^+)$ between the
data and fit(radians); the value plotted is that for each slice minus the
value from the fit.}
\label{figure6}
\end{figure}   

\begin{table}[htbp]
\begin{minipage}[]{0.9\linewidth}
\caption[]{Resonances fitted to the BES data and decay modes used
in the final fit.}
\label{table1}
\vspace{0.35cm}
\begin{tabular}{||cr|ll|cc|lr||}
\hline
      &                & \multicolumn{2}{c|}{Measured}          & \multicolumn{2}{c|}{Used in final fit} & decays & $\ell$ or $^{2S+1}\ell _2~$\\
\hline
$J^P$ & Resonance      & ~Mass               & ~~~$\Gamma$      & Mass  & $\Gamma$ & & \\
      &                & (MeV)               & (MeV)            & (MeV) & (MeV)    & & \\
\hline
$2^+$ & $f_2(1270)$ & $1280^{+35}_{-20} $ & ~~~~                & 1275  & 185 & $\rho\rho$        & $^1D_2~^3D_2$ \\
$0^-$ & ~~~~~~~~~~~ & $1440^{+20}_{-20} $ & $225^{+25}_{-20}$   & 1440  & 225 & $\rho\rho$        & $\ell = 1~~~~~~~~~~$ \\
$0^+$ & $f_0(1500)$ & $1505^{+15}_{-20} $ & $140^{+40}_{-30}$   & 1500  & 112 & $\sigma\sigma$    & $\ell = 0~~~~~~~~~~$ \\
$2^+$ & $f_2(1565)$ & $1505^{+60}_{-20} $ & $135^{+30}_{-25}$   & 1565  & 131 & $\rho\rho$        & $^5S_2~~~~~~~~~~~~$\\
$0^+$ & $f_0(1740)$ & $1740^{+30}_{-25} $ & $120^{+50}_{-40}$   & 1740  & 120 & $\sigma\sigma$    & $\ell = 0~~~~~~~~~~$ \\
$2^+$ & $f_2(1950)$ & $1940^{+50}_{-50} $ & $380^{+120}_{-~90}$ & 1940  & 380 & $f_2(1270)\sigma$ & $^5S_2~~~~~~~~~~~~$\\
      &             &                     &                     &       &     & $\sigma\sigma$    & $^1D_2~^3D_2$ \\
$0^+$ & $f_0(2100)$ & $2090^{+30}_{-30} $ & $330^{+100}_{-100}$ & 2104  & 215 & $\sigma\sigma$    & $\ell = 0~~~~~~~~~~$ \\
\hline
\end {tabular}
\end{minipage}
\end{table}

\begin{table}[htbp]
\begin{minipage}[]{0.9\linewidth}
\caption[]{Branching ratios for various final states in $J/\psi$ decays,
integrated up to $M(4\pi) = 2400 $ MeV. The first error is statistical  
and the second systematic.}
\label{table2}
\vspace{0.35cm}
\begin{tabular}{|ll|}
\hline
Process & Branching ratios \\
\hline
$Br(J/\psi\to\gamma f_2(1270)) \times Br(f_2(1270) \to\pi^{+}\pi^{-}\pi^{+}\pi^{-})$ & $(1.8\pm 0.2 \pm 0.6)\times 10^{-4}$ \\
$Br(J/\psi\to\gamma 0^-) ~~~~~~~~\times Br( 0^-\to\pi^{+}\pi^{-}\pi^{+}\pi^{-})$       & $(1.2\pm 0.1 \pm 0.4)\times 10^{-3}$ \\
$Br(J/\psi\to\gamma f_0(1500)) \times Br(f_0(1500) \to\pi^{+}\pi^{-}\pi^{+}\pi^{-})$ & $(3.1\pm 0.2 \pm 1.1)\times 10^{-4}$ \\
$Br(J/\psi\to\gamma f_2(1565)) \times Br(f_2(1565) \to\pi^{+}\pi^{-}\pi^{+}\pi^{-})$ & $(3.2\pm 0.2 \pm 1.1)\times 10^{-4}$ \\
$Br(J/\psi\to\gamma f_0(1740)) \times Br(f_0(1740) \to\pi^{+}\pi^{-}\pi^{+}\pi^{-})$ & $(3.1\pm 0.2 \pm 1.1)\times 10^{-4}$ \\
$Br(J/\psi\to\gamma f_2(1950)) \times Br(f_2(1950) \to\pi^{+}\pi^{-}\pi^{+}\pi^{-})$ & $(5.5\pm 0.3 \pm 1.9)\times 10^{-4}$ \\
$Br(J/\psi\to\gamma f_0(2100)) \times Br(f_0(2100) \to\pi^{+}\pi^{-}\pi^{+}\pi^{-})$ & $(5.1\pm 0.3 \pm 1.8)\times 10^{-4}$ \\
\hline
\end {tabular}
\end{minipage}
\end{table}

\begin{table}[htbp]
\begin{minipage}[]{0.9\linewidth}
\caption[]{Changes $\Delta S$ in log likelihood function S when various
components of the fit are dropped and remaining contributions are re-optimized.}
\label{table3}
\vspace{0.35cm}
\begin{tabular}{|cr|c|c|c|}
\hline
$J^P$ & Resonance   & $\Delta S$   & Freedom & Significance level \\
\hline
$2^+$ & $f_2(1270)  $ &  +18.67    &   5     & $~$5.0$\sigma$  \\
$0^+$ & $f_0(1500)  $ &  +21.79    &   4     & $>$6.0$\sigma$  \\
$2^+$ & $f_2(1565)  $ &  +16.32    &   4     & $~$4.8$\sigma$  \\
$0^+$ & $f_0(1740)  $ &  +17.73    &   4     & $~$5.1$\sigma$  \\
$2^+$ & $f_2(1950)  $ &  +37.30    &   7     & $>$6.0$\sigma$  \\
$0^+$ & $f_0(2100)  $ &  +23.58    &   4     & $>$6.0$\sigma$  \\
\hline
\end {tabular}
\end{minipage}
\end{table}

\begin{table}[htbp]
\begin{minipage}[]{0.9\linewidth}
\caption[]{Changes $\Delta S$ in log likelihood function when various
components are assigned a different $J^P$ and
all contributions are re-optimized.}
\label{table4}
\vspace{0.35cm}
\begin{tabular}{|cr|cccc|}
\hline
$J^P$ & Resonance      &   $0^-$   & $0^+$   & $2^+$     & $4^+$    \\
\hline
$2^+$ & $f_2(1270) $ &  +15.1    &  +13.2  &           &  +16.5   \\
$0^+$ & $f_0(1500) $ &  +18.0    &         &    +9.6   &  +17.2   \\
$2^+$ & $f_2(1565) $ &  +15.5    &  +16.0  &           &  +11.0   \\
$0^+$ & $f_0(1740) $ &  +16.8    &         &   +11.6   &  +15.7   \\
$2^+$ & $f_2(1950) $ &  +36.4    &  +28.2  &           &  +31.6   \\
$0^+$ & $f_0(2100) $ &  +13.2    &         &   +10.0   &  +18.1   \\
\hline
\end {tabular}
\end{minipage}
\end{table}

\end{center}

\end{document}